\documentclass{emulateapj}

\usepackage{graphicx}
\usepackage{epstopdf}
\usepackage{amsmath}
\usepackage{amssymb}
\usepackage{bm}
\usepackage{verbatim}
\usepackage{color}
\newcommand{\mbh}{$M_{\rm BH}$}
\newcommand{\mvir}{$M_{\rm vir}$}
\newcommand{\Hb}{H$\beta$}

\shorttitle{Dynamical Modeling of the Broad Line Region in Mrk 50}
\shortauthors{Pancoast et al.}

\begin{document}

\title{The Lick AGN Monitoring Project 2011: \\ Dynamical Modeling of the Broad Line Region in Mrk 50}

\author{
 Anna Pancoast\altaffilmark{1}, 
 Brendon J. Brewer\altaffilmark{1}, 
 Tommaso Treu\altaffilmark{1}, 
 Aaron J. Barth\altaffilmark{2}, 
 Vardha N. Bennert\altaffilmark{1,3}, 
 Gabriela Canalizo\altaffilmark{4}, 
 Alexei V. Filippenko\altaffilmark{5},
 Elinor L. Gates\altaffilmark{6}, 
 Jenny E. Greene\altaffilmark{7}, 
 Weidong Li\altaffilmark{5}, 
 Matthew A. Malkan\altaffilmark{8}, 
 David J. Sand\altaffilmark{1,9}, 
 Daniel Stern\altaffilmark{10},
 Jong-Hak Woo\altaffilmark{11},
 Roberto J. Assef\altaffilmark{10,12},
 Hyun-Jin Bae\altaffilmark{13},
 Tabitha Buehler\altaffilmark{14},
 S. Bradley Cenko\altaffilmark{5},
 Kelsey I. Clubb\altaffilmark{5},
 Michael C. Cooper\altaffilmark{2,15},
 Aleksandar M. Diamond-Stanic\altaffilmark{16,17},
 Kyle D. Hiner\altaffilmark{4}, 
 Sebastian F. H\"{o}nig\altaffilmark{1}, 
 Michael D. Joner\altaffilmark{14},
 Michael T. Kandrashoff\altaffilmark{5},
 C. David Laney\altaffilmark{14},
 Mariana S. Lazarova\altaffilmark{4}, 
 A. M. Nierenberg\altaffilmark{1}, 
 Dawoo Park\altaffilmark{11},
 Jeffrey M. Silverman\altaffilmark{5,18}, 
 Donghoon Son\altaffilmark{11},
 Alessandro Sonnenfeld\altaffilmark{1}, 
 Shawn J. Thorman\altaffilmark{2}, 
 Erik J. Tollerud\altaffilmark{2},
 Jonelle L. Walsh\altaffilmark{2,19}, 
and
 Richard Walters\altaffilmark{20} 
}

\altaffiltext{1}{Department of Physics, University of California,
Santa Barbara, CA 93106, USA; pancoast@physics.ucsb.edu}

\altaffiltext{2}{Department of Physics and Astronomy, 4129 Frederick
 Reines Hall, University of California, Irvine, CA, 92697-4575, USA}

\altaffiltext{3}{Physics Department, California Polytechnic State
University, San Luis Obispo, CA 93407, USA}

\altaffiltext{4}{Department of Physics and Astronomy, University of
 California, Riverside, CA 92521, USA}

\altaffiltext{5}{Department of Astronomy, University of California,
Berkeley, CA 94720-3411, USA}

\altaffiltext{6}{Lick Observatory, P. O. Box 85, Mount Hamilton, CA
95140, USA}

\altaffiltext{7}{Department of Astrophysical Sciences, Princeton
 University, Princeton, NJ 08544, USA}

\altaffiltext{8}{Department of Physics and Astronomy, University of
California, Los Angeles, CA 90095-1547, USA}

\altaffiltext{9}{Las Cumbres Observatory Global Telescope Network,
 6740 Cortona Drive, Suite 102, Santa Barbara, CA 93117, USA}

\altaffiltext{10}{Jet Propulsion Laboratory, California Institute of
Technology, 4800 Oak Grove Boulevard, Pasadena, CA 91109, USA}

\altaffiltext{11}{Astronomy Program, Department of Physics and
Astronomy, Seoul National University, Seoul 151-742, Republic of
Korea}

\altaffiltext{12}{NASA Postdoctoral Program Fellow}

\altaffiltext{13}{Department of Astronomy and Center for Galaxy
Evolution Research, Yonsei University, Seoul 120-749, Republic of
Korea}

\altaffiltext{14}{Department of Physics and Astronomy, N283 ESC,
Brigham Young University, Provo, UT 84602-4360, USA}

\altaffiltext{15}{Hubble Fellow}

\altaffiltext{16}{Southern California Center for Galaxy Evolution
Fellow}

\altaffiltext{17}{Center for Astrophysics and Space Sciences,
 University of California, San Diego, CA 92093-0424, USA}

\altaffiltext{18}{Physics Division, Lawrence Berkeley National Laboratory, 1 Cyclotron
Road, Berkeley, CA 94720, USA}

\altaffiltext{19}{Department of Astronomy, The University of Texas at
 Austin, Austin, TX 78712, USA}

\altaffiltext{20}{Caltech Optical Observatories, California Institute
of Technology, Pasadena, CA 91125, USA}

\begin{abstract}
We present dynamical modeling of the broad line region (BLR) in the
Seyfert 1 galaxy Mrk 50 using reverberation mapping data taken as
part of the Lick AGN Monitoring Project (LAMP) 2011.  We model the
reverberation mapping data directly, constraining the geometry and
kinematics of the BLR, as well as deriving a black hole mass estimate
that does not depend on a normalizing factor or virial coefficient.
We find that the geometry of the BLR in Mrk 50 is a nearly face-on
thick disk, with a mean radius of $9.6^{+1.2}_{-0.9}$ light days, a
width of the BLR of $6.9^{+1.2}_{-1.1}$ light days, and a disk opening
angle of $25 \pm 10$ degrees above the plane.  We also constrain the inclination
angle to be $9^{+7}_{-5}$ degrees, close to face-on.  Finally, the
black hole mass of Mrk 50 is inferred to be $\log_{10}($\mbh$/
M_\odot)=7.57^{+0.44}_{-0.27}$.  By comparison to the virial black hole mass
estimate from traditional reverberation mapping analysis, we find the
normalizing constant (virial coefficient) to be $\log_{10}f = 0.78 ^{+0.44}_{-0.27}$, 
consistent with the commonly adopted mean value of $0.74$ based on 
aligning the \mbh-$\sigma$* relation for AGN and quiescent galaxies.  While
our dynamical model includes the possibility of a net inflow or
outflow in the BLR, we cannot distinguish between these two scenarios.
\end{abstract}


\section{Introduction}

The standard model of active galactic nuclei \citep[AGNs;][]{antonucci93, urry95} explains their broad emission lines as being produced in a broad emission line region (BLR) situated on the order of light days from the black hole \citep{wandel99, kaspi00, bentz06}.  The distance of the BLR from the black hole can be measured using reverberation mapping, in which the average delay time is measured between a timeseries of the variable AGN continuum luminosity and a timeseries of the variable broad line emission \citep{blandford82, peterson93, peterson04}.  Standard reverberation mapping analysis also provides estimates of the black hole mass, \mbh, in AGNs from a normalized virial product.  The virial product, \mvir$ = f\,v^2\,c\,\tau/G$, is derived from the average light travel time lag of the BLR, $\tau$, and the typical velocity of the BLR gas, $v$, measured from the width of the broad lines.  The small sample of $\sim 50$ reverberation mapped AGNs is then used to determine single-epoch \mbh\ estimates for much larger samples of AGNs using the BLR-size to luminosity relation \citep{vestergaard06, mcgill08, vestergaard11}.  

However, there are certain limitations to the standard reverberation mapping techniques.  The object-to-object scatter of the normalization factor $f$ is believed to be of order $\sim0.4$ dex \citep{onken04, collin06, woo10, greene10, graham11} based on assuming the same \mbh-$\sigma$* relation \citep[e.g.][]{bennert11} as for quiescent galaxies.  It would be desirable to avoid this assumption and estimate \mbh\ from reverberation mapping data alone. 
The details of the BLR geometry and dynamics are also poorly constrained by standard reverberation mapping analysis.  Measuring the time lag as a function of line-of-sight velocity has shown that while some BLRs are consistent with virial motion in a Keplerian potential \citep{peterson99, bentz09, denney10}, some show suggestions of inflowing gas \citep{bentz10, denney10}.  In addition to the mean radius of the BLR as obtained in the standard analysis, we would like to constrain the overall geometry of the BLR in more detail.

Recent improvements in reverberation mapping data and analysis are starting to provide better constraints on the geometry and dynamics of the BLR.  Velocity-resolved transfer functions (VRTFs) have been measured using high-quality reverberation mapping data from the Lick AGN Monitoring Project in 2008 \citep[LAMP 2008,][]{walsh09, bentz09} and from the 2007 MDM Observatory reverberation mapping campaign \citep{denney10}, showing signatures consistent with inflow, outflow, and virialized motion for different AGNs.  However, a clear interpretation of VRTFs requires additional modeling steps, since they are functions of time lag instead of position within the BLR.  The traditional reverberation mapping analysis has also been recently improved by \citet{zu11}, who model the AGN continuum and line light curve using an implementation equivalent to Gaussian Processes \citep{kelly09,kozlowski10, macleod10, zu12}.  Members of our team have developed a method for determining the geometry and dynamics of the BLR by directly modeling reverberation mapping data \citep[][hereafter P11 and B11 respectively]{pancoast11, brewer11}, estimating the uncertainties in the framework of Bayesian statistics.  Our modeling method constrains \mbh\ without requiring a normalization constant $f$.  We also constrain the geometry of the BLR, its orientation with respect to the line of sight, and the possibility of net inflowing or outflowing gas in the BLR.  We have previously demonstrated our method on LAMP 2008 data for Arp 151 and estimated \mbh\ with smaller uncertainties than traditional reverberation mapping analysis (B11). 

What is now needed to make further progress is large samples of high quality
velocity resolved reverberation mapping data.  For this purpose we
carried out an 11-week spectroscopic observing campaign at Lick
Observatory, the Lick AGN Monitoring Project 2011.  The project focused on nearby AGNs with bright H$\beta$ lines, which are good candidates for dynamical
modeling.  Here we present the first results of dynamical modeling for the project, focusing on one of the most variable objects in the sample, Mrk
50.  The average time lag and virial \mbh\ estimates from
traditional reverberation mapping analysis are presented by
\citet{barth11}.  Here we present an alternative analysis based on our 
direct modeling technique.  The H$\beta$ and {\it V}-band continuum light
curve data are briefly described in Section~2, the dynamical model for
the BLR is described in Section~3, and our results and conclusions are
given in Section~4.

\begin{figure}[h!]
\begin{center}
\includegraphics[scale=0.43]{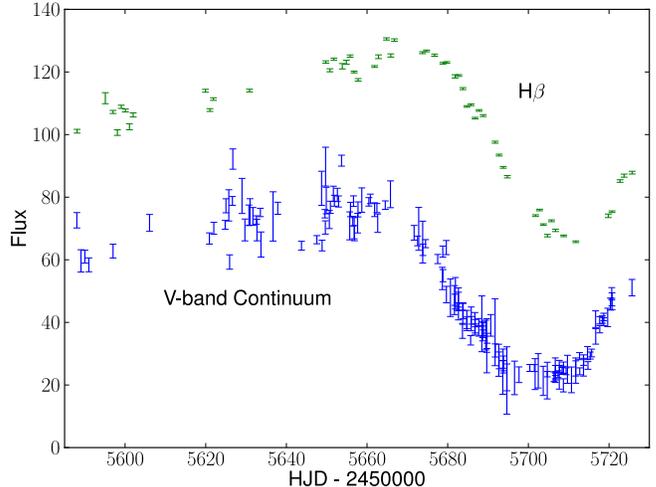}
\caption{The integrated H$\beta$ broad line and V-band continuum light curves.  The H$\beta$ light curve has flux units of $10^{-15}$\,erg\,cm$^{-2}$\,s$^{-1}$.  The V-band light curve is in arbitrary flux units.  
\label{fig_lc}}
\end{center}
\end{figure}

\begin{figure}[h!]
\begin{center}
\includegraphics[scale=0.4]{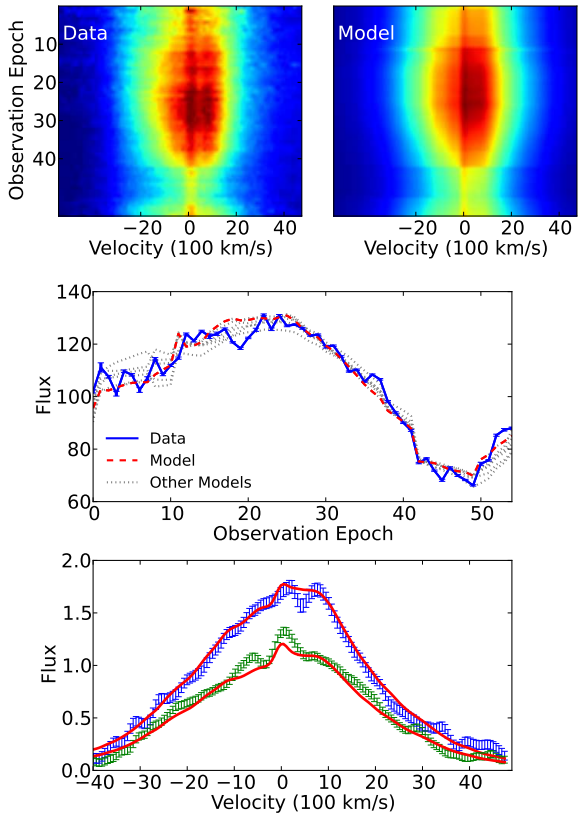}
\caption{Top: H$\beta$ spectra in velocity units for each epoch in the light curve for data, left panel, and model, right panel.  Dark red corresponds to the highest levels of flux and dark blue corresponds to the lowest levels, where the same color scale is used for the data and model.  Middle: integrated H$\beta$ flux for each epoch in the light curve for the data, blue solid line with errorbars, and model, red dashed line.  As an illustration of the range of solutions, we show light curves for five acceptable models as dotted gray lines.  For the correct time separation between light curve epochs, see Figure~\ref{fig_lc}.  The model is able to reproduce the major features of the data.  Bottom: two examples of H$\beta$ spectra fit by the model, with data shown by blue and green errorbars and model fits shown by red lines.  
\label{fig_rainbow}}
\end{center}
\end{figure}

\begin{figure}[t!]
\begin{center}
\includegraphics[scale=0.9]{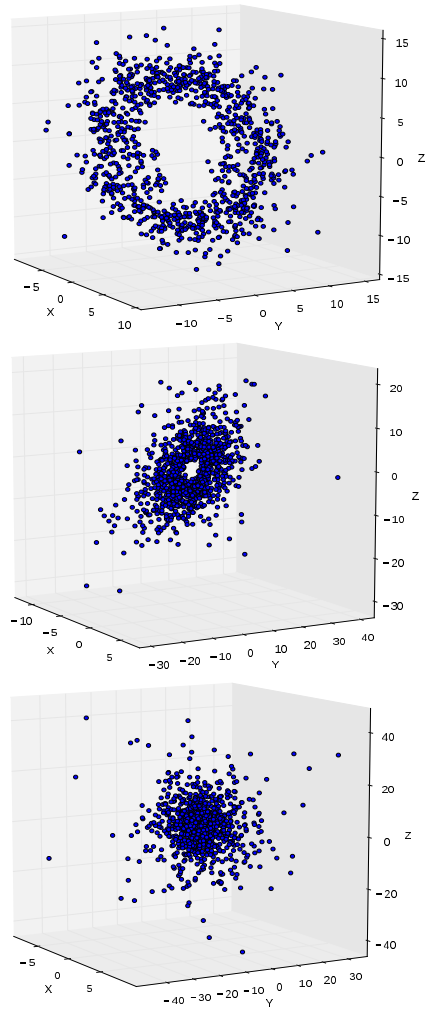}
\caption{Geometry of the BLR for three models, with the $x$, $y$, and $z$ axis scales in light days and the observer's line of sight along the $x$-axis.  The top panel BLR distribution is a close to face-on torus of clouds, the middle BLR distribution is a close to face-on disk of clouds similar to the geometry inferred for Mrk 50, and the bottom BLR distribution is a dense sphere of clouds.   
\label{fig_distr}}
\end{center}
\end{figure}

\begin{figure*}
\begin{center}
\includegraphics[scale=0.45]{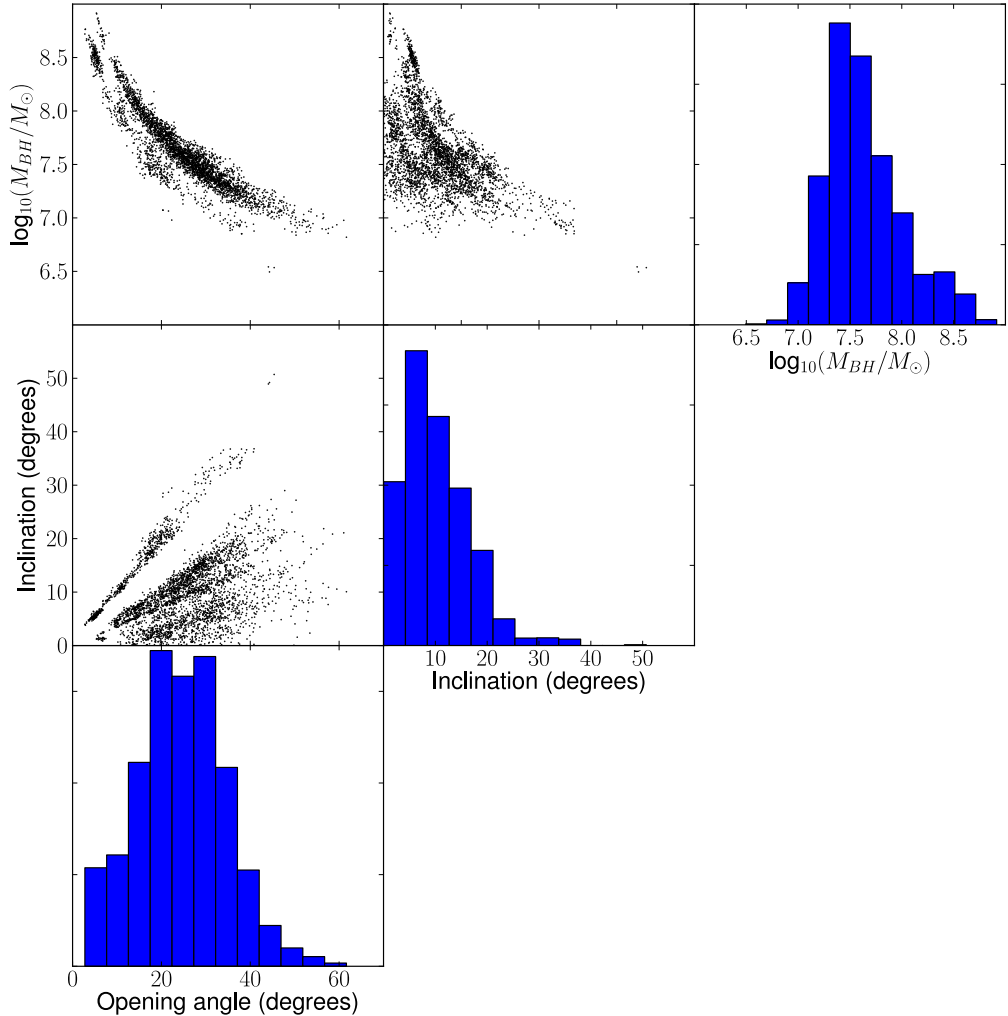}
\caption{Inferred posterior PDFs for model parameters, including \mbh, inclination angle (0 degrees is face-on), and opening angle of the BLR disk.  Joint posterior PDFs are also shown to illustrate the major degeneracies.  
\label{fig_cornerplot}}
\end{center}
\end{figure*}

\begin{figure}[h!]
\begin{center}
\includegraphics[scale=0.55]{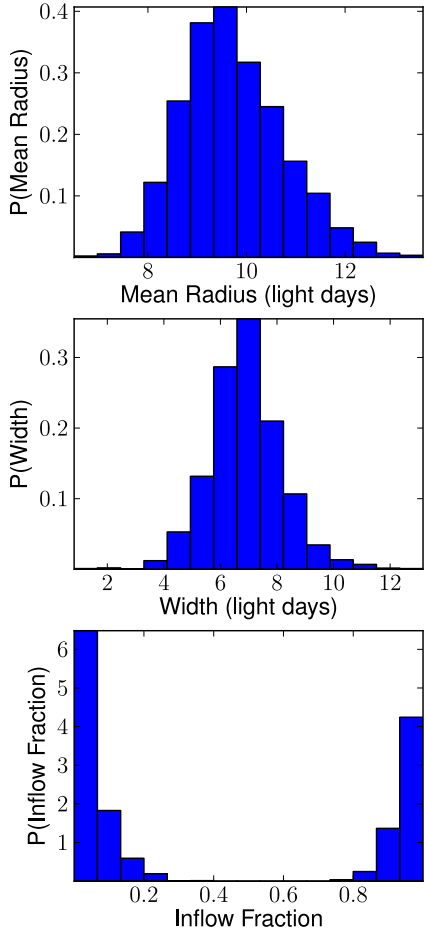}
\caption{Inferred posterior PDFs for model parameters, including the mean radius of the BLR, radial width of the BLR, and the inflow fraction of BLR gas.  
\label{fig_histplot}}
\end{center}
\end{figure}

\begin{figure}[h!]
\begin{center}
\includegraphics[scale=0.45]{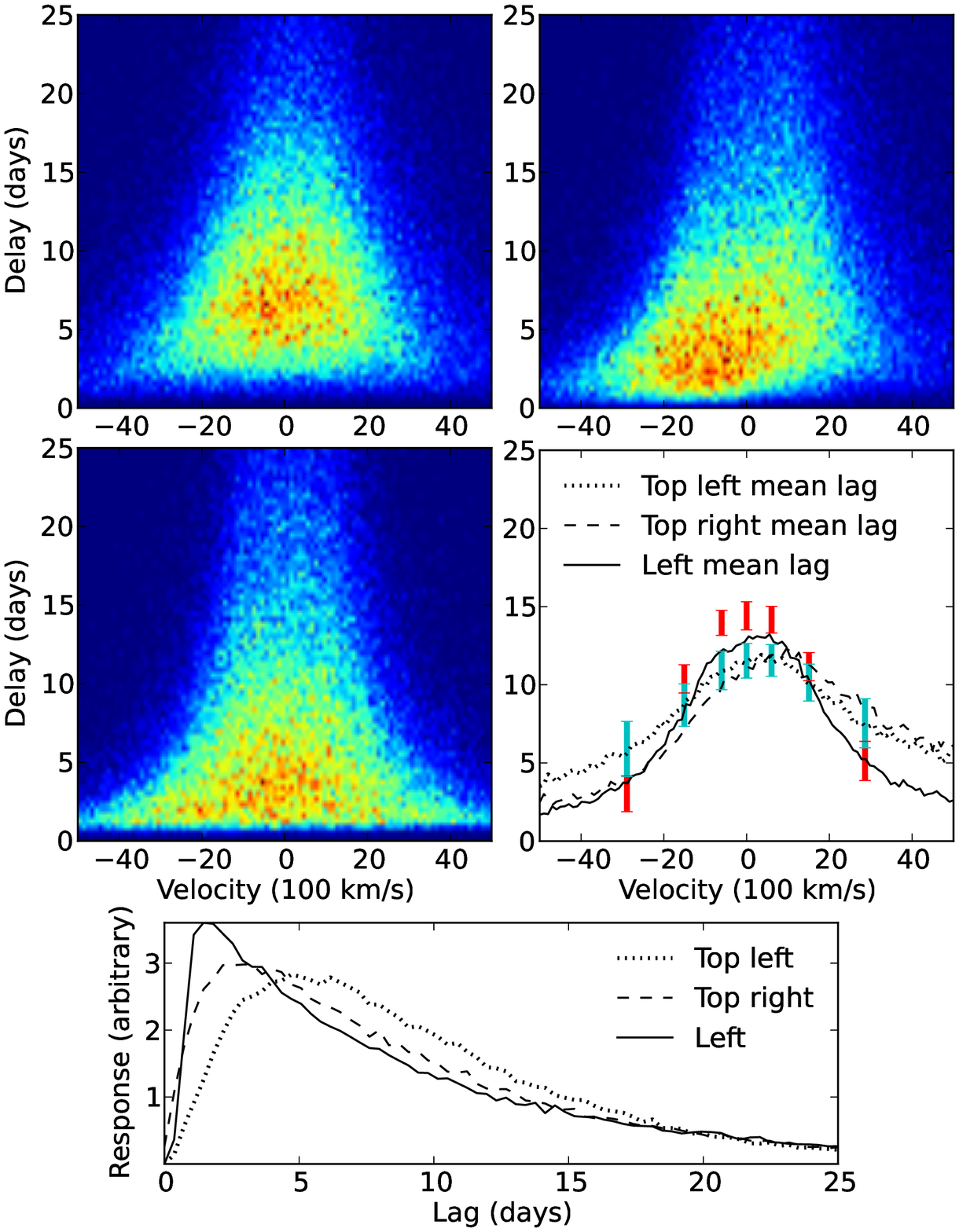}
\caption{Examples of acceptable transfer functions for Mrk 50.  The top two and middle left panels show examples of VRTFs drawn from the model parameter posterior PDFs, illustrating the range in inferred transfer function shape.  In the color-code of the VRTFs, red corresponds to the highest levels of response and dark blue corresponds to the lowest levels.  The middle right panel shows the mean lag for each of the VRTFs. 
The mean lag in seven velocity bins from \citet{barth11} are shown by red errorbars, which were measured by cross-correlation analysis.  We calculate the mean lag in the seven velocity bins of \citet{barth11} for $\sim 200$ VRTFs made using model parameters drawn randomly from their posterior PDFs, shown in light blue.  The bottom panel shows the velocity-integrated transfer functions for the VRTFs shown in the first three panels.
\label{fig_transfunct}}
\end{center}
\end{figure}

\section{Data}
We observed Mrk 50 in Spring 2011.  The data, shown in Figure\,\ref{fig_lc} and the top left panel of Figure\,\ref{fig_rainbow}, include a light curve of {\it V}-band continuum flux and a time series of the broad H$\beta$ line spectral profile.  More observational details, as well as details about the measurement of {\it V}-band and H$\beta$ light curves, are described by \citet{barth11}.  We model all 156 epochs of the {\it V}-band light curve and 55 of the 59 epochs of the \Hb\ line profile, ignoring those epochs with low $S/N$ or other problems.  The median $S/N$ for the H$\beta$ line profile throughout the campaign is 75 per pixel.  

AGN and stellar continuum lines can significantly alter the measured broad line widths in AGNs, affecting single epoch \mbh\ estimates \citep{denney09, park11}.  In order to reduce contamination from other lines when modeling the H$\beta$ line profiles, the Mrk 50 spectra have been fit with AGN and stellar continuum components and the \ion{He}{2} $\lambda 4686$ line just blueward of H$\beta$, and then these components were subtracted to yield the ``pure'' H$\beta$ profile \citep{barth11}.

\section{The Dynamical Model of the BLR}
We now give a brief description of our method for directly modeling
reverberation mapping data. The motivation for our approach is developed
in P11 and further implementation details are described in B11.  We model
the BLR as a large number of point-like clouds, each with a given
position and velocity. Several parameters describe the overall
spatial distribution of the clouds and the prescription for assigning velocities to the clouds, given their positions. Our goal is to estimate these parameters.

The continuum emission from the central
ionizing source is absorbed by these clouds and re-emitted as broad
line emission, allowing us to predict the line flux and shape as a
function of time, i.e. to produce mock data sets of the form shown in Figure~\ref{fig_rainbow}.

The full set of model parameters includes the
geometry and dynamics parameters for the BLR clouds corrected to the rest frame of Mrk 50, as well as a
continuous version of the continuum light curve, since the continuum
light curve must be evaluated at arbitrary times in order to compute mock data for comparison
with the actual data.  The observed continuum light curve is interpolated using Gaussian
Processes to create a continuous light curve and to account for the uncertainty in the interpolation.  Gaussian
Processes have been found to be a good model for larger samples of AGN
light curves \citep{kelly09, kozlowski10, macleod10, zu11, zu12}.  

The model for the BLR geometry is simple yet flexible, allowing for disk-like or
spherical geometries with asymmetric illumination of the gas.  Examples of
possible BLR geometries are shown in Figure~\ref{fig_distr}. The model for the spatial distribution of the BLR gas is first generated
from an axisymmetric 2-D configuration in the $x$-$y$ plane, with a parameterized radial profile.
The radius $r$ of a cloud from the origin is generated as follows. First, a variable
$g$ is drawn from a Gamma distribution with shape parameter $\alpha$ and scale parameter 1:
\begin{equation}
g \sim \Gamma\left(\alpha, 1\right)
\end{equation}
Then, the radius $r$ of the cloud is computed by applying the following linear
transformation to $g$:
\begin{equation}
r = F\mu + \frac{\mu\left(1 - F\right)}{\alpha}g \label{trans}
\end{equation}
The parameters $\left\{\mu, F, \alpha\right\}$ control the radial profile of the
BLR. $\mu$ is the overall mean radius of the BLR (this can be verified by taking
the expectation value of $r$ in Equation~\ref{trans}). The parameter $F \in [0, 1]$ allows 
for the possible existence of a hard lower limit $F\mu$ on radius, because there may be some radius interior
to which the BLR gas would all be ionized and thus unable to respond to changes in the continuum
emission \citep{korista04}. $\alpha$ controls the shape of the
gamma distribution: a value of $\alpha$ close to 1 imposes an exponential distribution (allowing for disk or ball configurations)
, whereas large values of $\alpha$ create a narrow normal distribution (allowing for shell or ring configurations). In the implementation,
and in the description of the same model in B11, we parameterise the shape by
$\beta = 1/\sqrt{\alpha}$ instead of $\alpha$ because $\beta$ has a simple interpretation as the standard
deviation of $g$ in units of its mean. The radial width of the BLR can be defined
as the standard deviation of $r$:
\begin{equation}
\sigma_r = \mu\beta(1-F).
\end{equation}

In order to assign velocities to the BLR gas clouds, the model uses probabilistic perturbations about circular orbits.  The solution for the radial velocity of a BLR cloud given its position $r$, energy $E$, and angular momentum $L$ is:
\begin{equation}
\label{eqn_rdot}
 v_r = \pm \sqrt{2 \left( E+\frac{GM_{\rm BH}}{r} \right) -\frac{L^2}{r^2}}.
\end{equation}
If we wished to impose circular orbits, the values for $E$ and $L$ would be fully determined by the radius $r$ of the cloud:
\begin{eqnarray}
 E_{\rm{circ}} &=& -\frac{1}{2}\frac{GM_{\rm BH}}{r}    \\
 L_{\rm{circ}} &=& \pm r \sqrt{2\left( E + \frac{GM_{\rm BH}}{r} \right)}.
\end{eqnarray}
To obtain elliptical orbits, we generate values for $E$ and $L$ probabilistically, given $r$. The probability distributions for energy and angular momentum are parameterized by the parameter $\lambda$ and are given by:
\begin{eqnarray}
 E = \left( \frac{1}{1+ {\rm exp}(-\chi)} \right) E_{\rm circ}  \\  
 p(L) \propto {\rm exp} \left( \frac{|L|}{\lambda}  \right)    
\end{eqnarray}
where $\chi \sim \mathcal{N}(0,\lambda^2)$ and $|L| < |L_{\rm circ}|$.  For $\lambda \rightarrow 0$ we recover circular orbits and increasing $\lambda$ creates more elliptical orbits.  Since there are two solutions for the sign of $v_r$, the model also includes a parameter for the fraction of outflowing vs inflowing gas.  The inflowing and outflowing gas is bound to the gravitational potential of the black hole, but an inequality in the fraction of inflowing and outflowing gas has the desired effect of modeling asymmetries in the H$\beta$ spectral line profile as observed in Arp 151 (B11) when an asymmetric illumination model is included.

Once a 2-D configuration of clouds in the $x$-$y$ plane has been generated, and velocities have been assigned to the clouds,
rotations are applied to ``puff up'' the 2-D configuration into a 3-D configuration. We first rotate the cloud positions about
the $y$ axis by a small random angle; the typical size of these angles
determines the opening angle of the cloud distribution.  The opening angle is defined as the angle above the midplane of the disk or sphere. We then rotate around the $z$ axis by random angles to restore the axisymmetry of the model. Finally, we rotate
again about the $y$ axis, by the inclination angle (common to all of the clouds) to model the
inclination of the system with respect to the line of sight.  The inclination angle is defined so that zero degrees corresponds to a face-on configuration and 90 degrees corresponds to an edge-on configuration.

In order to produce asymmetric broad line
profiles, we include a simple prescription for asymmetric illumination
of the BLR clouds. We assign a weight $w$ to each cloud, given by $w = 0.5 + \kappa \cos \phi$, where $\phi$ is the azimuthal position of the cloud in spherical polar coordinates.  The parameter $\kappa$ ranges from $0.5$, corresponding to illuminating the near side of the BLR, to $-0.5$, corresponding to illuminating the far side of the BLR.
Physically, the near side of the BLR could be preferentially illuminated if the
far side of the BLR were obscured by gas, and the far side of the BLR
could be preferentially illuminated if the BLR clouds only reradiate
the continuum emission towards the central ionizing source due to
self-shielding within the cloud.  Inflowing gas with the near side of the BLR illuminated can, in principle, be distinguished from outflowing gas with the far side of the BLR illuminated by the VRTF, since the lags for these two cases are different.  

In addition, we allow for a scaling factor to describe the percentage variability of the emission line compared to that of the continuum.  
While for Arp 151 the variability of the continuum was approximately equal to that of the H$\beta$ flux, in the case of Mrk 50 we find that the continuum variability is less than that of the line.  This is consistent with the amplitude of variability of the ionizing continuum responsible for \Hb\ being larger than that of the V-band \citep[][and references therein]{meusinger11}.

Once the dynamical model has been defined, we are able to compute simulated data that are then blurred with a Gaussian kernel to model the instrumental resolution. The simulated data are then compared with
the actual data. For the likelihood function, we use the standard Gaussian assumption:
\begin{equation}
P(\mathrm{data}|\mathrm{model}) \propto \exp\left[-\frac{1}{2}\chi^2(\mathrm{model}, \mathrm{data})\right]
\end{equation}

With the likelihood function defined, the modeling problem is
reduced to computing the inferences on all of the model parameters.
The likelihood function, $P(\mathrm{data}|\mathrm{model})$, is
combined with the prior distribution for the parameters using Bayes'
Theorem: $P(\mathrm{model} | \mathrm{data}) \propto P(\mathrm{model}) \times P(\mathrm{data}
| \mathrm{model})$.  The posterior probability distribution for the parameters
is sampled using the Diffusive Nested Sampling algorithm \citep{brewer09}.  
Nested Sampling algorithms initially sample the prior distribution,
and subsequently create and sample more constrained distributions,
climbing higher in likelihood. In the specific case of Diffusive
Nested Sampling, uphill and downhill moves are allowed, allowing the
exploring particles to return to the prior, take large steps, and then
climb the likelihood function again.
We assigned uniform priors to most parameters except for the mean radius and \mbh, which have log uniform priors to describe initial uncertainty about the order of magnitude of the parameter.

By computational necessity, our model is relatively simple. While it is still rather flexible and can reproduce the large scale
features of the reverberation mapping data, it is unable to model every detail of the H$\beta$ light curve.  The large scale features of the 
variability in the H$\beta$ light curve are well-modeled, for example,
but the small epoch-to-epoch fluctuations in the light curve are not (see Figure~\ref{fig_rainbow}). In addition, the errorbars reported on the data are very small, and our model is not able to fit the data set to within these small error bars (i.e. we cannot achieve reduced $\chi^2 \sim 1$). If we did not take this into account our uncertainties would be unrealistically small. This issue is a generic feature of the fitting of simply parameterized models to informative data sets, and will be discussed in depth in a forthcoming contribution (Brewer et al, in preparation). In order to account for this effect and to obtain realistic and conservative uncertainties, we explore the effect of inflating the errorbars on the spectrum data by a factor $H$, or equivalently, choosing to form the posterior distribution from different chunks of the Nested Sampling run (i.e. different ranges of allowed likelihood values). For each value of $H$ tested, we inspect the posterior distribution over simulated data (top right panel in Figure~\ref{fig_rainbow}) to ensure that the major features of the data are reproduced. We find that, as long as $H$ is low enough that the models fit the major features in the data, the resulting posterior distributions on the parameters are insensitive to the exact choice of the value for $H$.

\section{Results and Conclusions}
Our inferred geometry and dynamics parameters of the BLR in Mrk 50 are
shown in Figures~\ref{fig_cornerplot} and \ref{fig_histplot}.  The shape of the BLR gas radial
profile is constrained to be closer to exponential ($\alpha \lesssim 1$), with a mean radius
of $\mu = 9.6^{+1.2}_{-0.9}$ light days and a width 
of $\sigma_r = 6.9^{+1.2}_{-1.1}$
light days (the uncertainties quoted are symmetric 68\% confidence limits). 
Even though the mean radius is not simply $c$ times the mean lag in the general
asymmetric case, we expect our mean radius to roughly correspond to the
lag measurements using cross-correlation analysis by
\citet{barth11}, which are
$\tau_{\rm peak} = 9.75^{+0.50}_{-1.00}$ and $\tau_{\rm cen} = 10.64^{+0.82}_{-0.93}$
light days. Our mean radius agrees more closely with $\tau_{\rm peak}$,
although $\tau_{\rm cen}$ is more commonly used for black hole mass estimation.
We infer the inner radius of the BLR distribution to be
$F\mu = 2.0^{+1.3}_{-1.1}$ light days.
The opening angle of the BLR disk, defined between 0-90 degrees, is $25 \pm 10$ degrees, closer
to a thin disk than to a sphere.  The inclination angle of the thick
BLR disk with respect to the line of sight is constrained to be
$9^{+7}_{-5}$ degrees, closer to face-on, consistent with the
standard model of broad line AGNs \citep{antonucci93, urry95}.

The dynamical modeling results constrain Mrk 50 to have $39\%$ probability
of net inflowing gas and $61\%$ probability of net outflowing gas,
with equal amounts of inflowing and outflowing gas ruled out (inflow
fraction $= 0.5$), as shown in Figure~\ref{fig_histplot}.  
This result suggests only a slight preference for outflow while the need for either outflow or inflow is quite robust, suggesting that a more physical model for inflow and outflow is needed in order to distinguish between them for the case of Mrk 50.
  Equal amounts of inflowing and outflowing gas are ruled out because net
inflowing or outflowing gas, along with the illumination model, creates the asymmetry in the \Hb\ line
profile observed in the data.  

In addition to constraining the geometry of the BLR, our
dynamical model also places an independent estimate on \mbh, inferred
to be $\log_{10}($\mbh$/M_\odot) = 7.57^{+0.44}_{-0.27}$.  Part of the uncertainty in this estimate comes from the range in possible \mbh\ values at nearly face-on inclinations (close to zero degrees), as shown in Figure~\ref{fig_cornerplot}.  Recent
cross-correlation reverberation mapping results quote statistical uncertainties of
the order of 0.15 dex \citep{bentz09, denney10, barth11a, barth11},
but this neglects the uncertainty in the normalization factor, $f$,
that is believed to have an object to object scatter of
0.44 dex \citep{woo10, greene10b}. Thus, our uncertainty in \mbh\ for Mrk 50 is
smaller that achieved by traditional reverberation
mapping estimates.  Our independent measurement of
\mbh\ can be used to estimate the appropriate value of $f$ for Mrk
50 by comparing it to the virial estimate by
\citet{barth11}, \mvir\,=\,$ f\,v^2\,c\,\tau/G$, where $\tau$ and $v$
are obtained from the cross-correlation of the continuum and broad
line light curves and from the width of the broad line, respectively.
We find $\log_{10}f =0.78^{+0.44}_{-0.27}$, which agrees to within the
errors with the commonly used mean values of $\log_{10} \left \langle f \right \rangle
=0.74^{+0.12}_{-0.17}$ from \citet{onken04}, $\log_{10} \left \langle f \right \rangle
=0.72^{+0.09}_{-0.10}$ from \citet{woo10}, and $\log_{10} \left \langle f \right \rangle
=0.45^{+0.09}_{-0.09}$ from \citet{graham11}.  We have used $\left \langle f \right \rangle$ to denote a normalization factor derived from large samples of reverberation mapped AGN \mbh\ estimates as distinct from the $f$ value we measure individually for Mrk 50.  A sample of 10 independent black hole mass and $f$ measurements with comparable uncertainties to Mrk 50 and Arp 151 would allow us to calculate a mean $f$ value to $\sim 0.3/\sqrt{10} \simeq 0.1$ dex uncertainty and to distinguish between these commonly used mean values.

An additional interesting feature of Figure~\ref{fig_cornerplot} is the complex structure in the joint posterior distribution for the inclination angle and opening angle, a feature that was not seen in Arp 151. The joint posterior appears to have two distinct families of solutions, although one has almost four times as much weight as the other.  In an attempt to understand the origin of this structure, we separated the posterior samples in the two modes in order to test whether they are correlated with any other parameters (such as the inflow fraction), however we were unable to find any such correlations. Future improvements to the flexibility and realism of the model may enable us to rule out one of these modes, and hence constrain the parameters more tightly and further reduce the uncertainties.

While \mbh\ is well constrained, there are many ways to successfully
model the large-scale structure of the reverberation mapping
data. This is illustrated by the degeneracies in the posterior
distributions plotted in Figure~\ref{fig_cornerplot}.  The quality of
the model fits to the data are illustrated in
Figure~\ref{fig_rainbow}, including six model integrated H$\beta$ flux
light curves, an example of a model dataset of spectra for each epoch
in the light curve, and two data spectra with the model spectra
overplotted.  The smoothness of the models compared to the data is
illustrated in the spectral datasets of the data and model shown in
the top panels of Figure~\ref{fig_rainbow}.  The Mrk 50 H$\beta$
spectral profile did not change in shape drastically over the course
of the LAMP 2011 reverberation mapping campaign, and the model
spectral profile is likewise very similar for all epochs.  Even though
the shape of the individual spectral profiles can be well-modeled,
more sophisticated models will be required to match the detail of the
small-scale variability of the integrated H$\beta$ data light curve.

Note that the uncertainties quoted throughout this paper are determined from a Monte Carlo method, and are therefore subject to error themselves. As we are interested in reducing the uncertainties on black hole mass estimates from reverberation mapping data, it is important to quantify the uncertainty on the uncertainty. To investigate this, we estimated the effective number of independent samples produced by our Diffusive Nested Sampling runs, by counting the number of times the exploring particles returned to the prior (allowing large steps to be taken) before climbing the likelihood peak again. Our effective number of independent samples was found to be $\sim 180$. We then generated samples of size 180 from our full posterior sample, and determined the scatter in the resulting $\log_{10}(M_{{\rm BH}})$ uncertainties to be 0.02. Thus, the uncertainty on the black hole mass for Mrk50 is $^{+0.44}_{-0.27} \pm 0.02$ dex.

Previous attempts to understand the geometry and dynamics of the BLR
have focused on reconstructing the VRTF
\citep{kollatschny02, bentz10, denney10}. In the interests of 
comparing future transfer function studies to our physically motivated model of
the BLR, we show three inferred VRTFs
for Mrk 50 in Figure~\ref{fig_transfunct}. These three transfer functions were chosen out of the many inferred
possible models for Mrk 50 to show some of the variety in allowed transfer
function shapes.  The top left VRTF has a fairly typical shape and level of asymmetry, while 
the top right VRTF is more asymmetric than average.  One measurement of the VRTF asymmetry is 
to compare the integral of the mean lag per velocity bin on either side of line center, 
corresponding to the zero velocity point in the middle right panel of Figure~\ref{fig_transfunct}.  
By this measurement of asymmetry, 43\% of the possible models inferred for Mrk 50 have VRTFs that 
are less asymmetric than the top left VRTF, while only 8\% of the possible models have VRTFs that 
are more asymmetric than the top right VRTF.   The middle left transfer function illustrates the extent to which our inferred model for Mrk 50 can agree with the velocity-resolved cross-correlation measurements by \citet{barth11}, shown by red errorbars in the middle right panel of Figure~\ref{fig_transfunct}.  This VRTF has the smallest $\chi^2$ distance from the cross-correlation measurements by \citet{barth11} and models with this level of agreement (or better) have a probability of $\sim 0.3\%$.
The average shape of the VRTF is also shown in
Figure~\ref{fig_transfunct}, with the same velocity bins as used by
\citet{barth11} for their cross-correlation based measurement.  
This average VRTF is fairly symmetric, but the higher velocity bins have larger errorbars as a result of averaging over transfer functions that have asymmetries from either net inflowing or outflowing gas (see the dashed line in the middle right panel of Figure~\ref{fig_transfunct}).

Note that the average VRTF we infer and the 
results obtained from cross-correlation measurements by \citet{barth11} do not
all agree to within the 1-$\sigma$ error bars.  In order 
to understand the differences between the time-lags as measured in our dynamical 
model and those measured through the cross-correlation procedure, we consider the 
ideal continuous noise-free case.  In this case, the cross-correlation function (CCF) between 
the line and continuum light curves is the transfer function convolved with the 
autocorrelation function (ACF) of the continuum light curve, which is the CCF of 
the light curve with itself.  While the ACF is symmetric, the transfer function 
may be asymmetric, as we find for Mrk 50, so the CCF may also be asymmetric.  One 
measurement of the cross-correlation time-lag often used to measure black hole 
mass is the CCF-weighted mean lag, $\tau_{\rm cen}$, which is by definition 
affected by the asymmetry in the CCF.  Therefore, in the case of asymmetric 
transfer functions, $\tau_{\rm cen}$ may not correspond to the mean lag of our 
dynamical model of the BLR.  For the non-ideal case, a direct comparison between 
cross-correlation measurements and the results of our dynamical modeling approach 
is not straightforward, since the peak (or mean) of the CCF does not measure the 
true mean lag but only a noisy version of the convolution between the ACF and the 
transfer function.  

We explored this issue by running the cross-correlation
technique as implemented by \citet{barth11} on light curves generated
by models drawn from the posterior PDF for Mrk 50. For simplicity we
considered noise-free light curves sampled in the same way as our
data. We find that the peak and CCF-weighted mean ($\tau_{\rm peak}$ and 
$\tau_{\rm cen}$) of the CCF can be
systematically off by $\sim1-2$ light days with respect to the true 
mean lag of the model. The amount of the offset varies as a
function of the actual shape of the transfer function as well as the
details of the implementation of the cross-correlation
algorithm. Thus, it is not surprising that we find systematic
differences of this order between our estimates of the mean
lag and $\tau_{\rm cen}$. Clarifying and
quantifying systematically the relationship between these two
approaches as a function of BLR structure and data quality is an
important topic that goes beyond the scope of this paper and is left
for future work. 

In conclusion, the analysis presented here provides new and unique
insights into the geometry and kinematics of the BLR, and a \mbh\
estimate that is competitive with the most accurate methods.  However,
since our modeling uncertainties are greater than data uncertainties,
more physical models that take into account the complex processes
occurring in the BLR should allow for even better constraints.  In the
future, we plan to develop such models and apply them to large samples
of reverberation mapping data.

\acknowledgments
We thank the Lick Observatory staff for their exceptional support during our observing campaign. In addition, we thank Brandon Kelly for suggesting changes to our code that yielded significant improvements. The referee also provided valuable feedback that enabled us to improve the paper. The Lick AGN Monitoring Project 2011 is supported by NSF grants AST-1107812, 1107865, 1108665, and 1108835. The West Mountain Observatory is supported by NSF grant AST-0618209.  AP acknowledges support from the NSF through the Graduate Research Fellowship Program. BJB and TT acknowledge support from the Packard Foundation through a Packard Fellowship to TT.  AD acknowledges support from the Southern California Center for Galaxy Evolution, a multi-campus research program funded by the University of California Office of Research.  AVF and his group at UC Berkeley acknowledge generous financial assistance from Gary \& Cynthia Bengier, the Richard \& Rhoda Goldman Fund, NASA/{\it Swift} grants NNX10AI21G and GO-7100028, the TABASGO Foundation, and NSF grant AST-0908886.  SH acknowledges support by Deutsche Forschungsgemeinschaft (DFG) in the framework of a research fellowship (``Auslandsstipendium'').  The work of DS and RA was carried out at the Jet Propulsion Laboratory, California Institute of Technology, under a contract with NASA. JHW acknowledges support by Basic Science Research Program through the National Research Foundation of Korea funded by the Ministry of Education, Science and Technology (2010-0021558).

\end{document}